\documentclass[aps,12pt,prepreint,prc,notitlepage,floatfix,nofootinbib,superscriptaddress,amsmath,amssymb,longbibliography]{revtex4-1}

\usepackage[utf8]{inputenc}
\usepackage[T1]{fontenc}
\usepackage{lmodern}

\usepackage{mathrsfs}
\usepackage{ulem}
\usepackage{graphicx}
\usepackage{epsfig}
\usepackage{bm}
\usepackage{color}
\usepackage{float}
\usepackage{dcolumn}
\usepackage{multirow}
\usepackage{xcolor}
\usepackage{hyperref}
\hypersetup{colorlinks,
            linkcolor={red!50!black},
            citecolor={blue!50!black},
            urlcolor={blue!80!black}
}
\usepackage{color}
\usepackage{braket}
\usepackage{amsmath}

\begin{document}

\title{Renormalization of a Finite Range Inverse Cube Potential}


\author{D. Odell}

\affiliation{Department of Physics and Astronomy, University of
  Tennessee, Knoxville, Tennessee 37996, USA}

\author{A. Deltuva}
\affiliation{Institute of Theoretical Physics and Astronomy,
Vilnius University, Saul\.etekio al. 3, LT-10257 Vilnius, Lithuania}

\author{J. Bonilla}
\affiliation{Department of Physics and Astronomy, University of
  Tennessee, Knoxville, Tennessee 37996, USA}

\author{L. Platter}
\affiliation{Department of Physics and Astronomy, University of
  Tennessee, Knoxville, Tennessee 37996, USA}
\affiliation{Physics Division, Oak Ridge National Laboratory, Oak
  Ridge, Tennessee 37831, USA}

\begin{abstract}
  We study the regularization and renormalization of a finite range
  inverse cube potential in the two- and three-body
  sectors. Specifically, we compare and contrast three different
  regulation schemes frequently used to study few-body systems as well
as the associated renormalization group (RG) flows. We also
  calculate bound state and scattering observables over a wide range
  of cutoffs, demonstrating the sufficiency of a two-body contact
  interaction to renormalize two- and three-body observables. We
  supplement these plots with quantified analyses of the observables'
  residual cutoff dependence.

\end{abstract}
\smallskip
\maketitle

\section{Introduction}\label{intro}
Effective field theories (EFTs) have become a standard tool in nuclear few-body
physics to construct the interactions between the considered degrees of
freedom~\cite{Epelbaum:2008ga,Hammer:2012id}. For example, chiral effective
theory is a low-energy expansion of the nucleon-nucleon ($NN$) interaction that
employs only nucleons and pions as degrees of freedom and that uses the pion
mass $m_\pi$ (or a small momentum) over a large scale $\Lambda$ that can be
associated with the lightest degree of freedom not included in the EFT (e.g.~the
$\rho$-meson). This framework is then used to derive the nuclear Hamiltonian in
a systematic low-energy expansion. The resulting potential has been used
extensively in few-nucleon studies and ab initio nuclear structure calculations.
It was pointed out that the most singular piece of the one-pion exchange (OPE)
in the deuteron channel is an inverse cube
potential~\cite{Sprung:1994,PavonValderrama:2005gu}.  The renormalization of
this leading order (LO) potential has been studied repeatedly in the two- and
three-nucleon
sector~\cite{PavonValderrama:2004nb,Nogga:2005hy,Birse:2005um,Long:2011xw,Song:2016ale}.
Here, we study the renormalization of the finite range inverse cube
potential (FRIC) in the much simpler three-boson system thereby
removing the complications due to
the spin-dependent tensor force. In particular, we examine whether the
three-body system with pairwise inverse cube interactions requires a three-body
counterterm for renormalization, and whether residual cutoff corrections can be
used as a reliable tool to build a power counting scheme as suggested in
Ref.~\cite{Griesshammer:2015osb}. We note that there is also interest in atomic
physics regarding the inverse cube interaction. However, most attention is
focused on the low-energy properties in the \textit{infinite} range
limit~\cite{Mueller:2013,Gao:1999}.

Since the residual cutoff dependence to some extent can be influenced
by the chosen regularization scheme, we carry out this analysis
for various schemes that are currently used by the
community. Specifically, we consider a \textit{local}
regularization scheme~\cite{Gezerlis:2013ipa} that cuts off the
potential in coordinate space at a small distance $R$, a non-local
regularization scheme~\cite{Epelbaum:2008ga} that cuts off the high momenta
in the momentum space form of the two-body interaction $V(p,p')$
separately, and a semi-local regularization scheme~\cite{Epelbaum2015}
that applies these strategies separately to the long-range inverse
cube part of the interaction and the short-distance regulator.

These different regularization schemes have different advantages for
different methods that are used to diagonalize the nuclear
Hamiltonian. For example, local interactions are commonly used in
quantum Monte Carlo calculations, though progress has been made
including nonlocal interactions
(e.g.~\cite{Roggero:2014kea,Lynn:2012}). However, while these have
been used extensively in the literature, a detailed comparison of
these approaches is missing.

We find that the regularization schemes analyzed can be used to obtain
regulator-independent results at large cutoffs. We find however that the regulator
dependence of the short-distance counterterm is different for the
regulation schemes we apply.  In agreement with findings in the
three-nucleon sector\cite{Nogga:2005hy, Song:2016ale}, we find that
three-body observables are completely renormalized without the
inclusion of an additional three-body counterterm. However, an
analysis of the cutoff dependence of three-body observables shows also
that observables converge more slowly than expected from previous
studies of the three-nucleon sector~\cite{Song:2016ale}.

In Sec.~\ref{sec:theory}, we discuss the regularization schemes as well as the
renormalization and calculation of observables. In
Sec.~\ref{sec:results}, we present the results obtained for the two- and
three-boson system as well as quantitative analyses of the remaining cutoff
corrections. We conclude with a summary and an outlook.

\section{Theory}
\label{sec:theory}

In the following subsections, we describe the interaction that is used
in this work, how it is regulated, and how it is renormalized. We
comment also briefly on technical details such as the normalization of
states and the calculation of observables through the Schr\"odinger,
Lippmann-Schwinger, and Faddeev equations.

The non-regulated and singular potential $V_S$ that we consider is a FRIC
potential of the form
\begin{equation}
	\label{eq:fric_pot}
  V_{\rm{S}}(r) = -C_3 \frac{e^{-m_\pi r}}{r^3}~.
\end{equation}
We choose $m_\pi = 138$ MeV and $C_3 = 0.8$ fm$^2$ such that a
deuteron-like state ($B_2=2.2$ MeV) exists when we regulate the
potential at $\sim 1~\textrm{fm}$. This potential has to be regulated
at short distances and observables will depend strongly on the
regularization scale as the interaction is too singular~\cite{Frank:1971xx}.
Below we display how a (\textit{smeared out}) short-distance counterterm can be
introduced to address this problem.

We perform our calculations in momentum space, and we Fourier transform the
interaction $V$ and carry out a partial-wave projection
\begin{equation}
  \label{eq:ft_pwp}
  \tilde{V}_{l}(p,k) \equiv FT\left[V(r)\right] =
  \frac{2}{\pi}\int_0^\infty drr^2 j_l(pr) V(r) j_{l}(kr)~,
\end{equation}
where $j_l(z)$ are the spherical Bessel functions of order $l$.

\subsection{Regulator Formulations}\label{sec:regs}

\subsubsection{Local Regulation}\label{sec:local_reg}

For a local, singular potential, $V_S(r)$, we have
implemented three different forms of regulation: local, semi-local, and
nonlocal. The locally regulated potential has the form
\begin{equation}
  \label{eq:local}
  V(r) = \rho(r;R)V_S(r) + g(R)\chi(r;R)~,
\end{equation}
where $\rho(r;R)$ is an arbitrary function that minimally fulfills two requirements.
First, it must overcome $V_S(r)$ in the $r\rightarrow 0$ limit such that the
product $\rho(r;R)V_S(r)$ is finite.
Second, in the limit of $r\rightarrow\infty$, $\rho(r;R)$ must go to one.
For the locally regulated case we use
\begin{equation}
  \label{eq:local_reg}
  \rho(r;R) = {\left(1-e^{-{(r/R)}^2}\right)}^4~,
\end{equation}
where $R$ is the range at which the characteristic behavior of $V_S(r)$ is cut
off. The counterterm
\begin{equation}
  \label{eq:local_cterm}
  g(R)\chi(r;R)~,
\end{equation}
has two components. The first, $g(R)$ is an $R$-dependent coupling
strength. We tune this parameter to match some low-energy, two-body
observable such as the two-body binding energy. The second,
$\chi(r;R)$, is a contact-like interaction or a \textit{smeared} $\delta$
function such that
\begin{equation}
	\lim_{R\rightarrow 0}\chi(r;R) \sim \delta(r)~.
\end{equation}
For the locally regulated case we use
\begin{equation}
  \label{eq:locally_xterm}
  \chi(r;R) = e^{-{(r/R)}^3}~.
\end{equation}
We discuss below that the RG flow of the locally-regulated
counterterm strength, $g(R)$, contains multiple
branches~\cite{Beane:2000wh}. To ensure consistency between our
results and others', we also implement a semi-local regulation
scheme.

\subsubsection{Semi-Local Regulation}\label{sec:semi_local_reg}

The difference between local regulation and semi-local regulation
lies in the definition of the counterterm. In Eq.~\eqref{eq:local}
we defined the counterterm in coordinate space. This counterterm,
that regulates the relative distance in the two-body system and
thereby the momentum exchange, has multiple solutions (provided the
short-distance cutoff is small enough) for which the two-body binding
energy $B_2$ is reproduced.

If we instead define the counterterm in momentum space as
\begin{equation}
  \label{eq:momspace_cterm}
  g(R)\tilde{\chi}(p;R)\tilde{\chi}(k;R)~,
\end{equation}
such that, by itself, only permits one state, we obtain a unique RG flow.
The full potential in momentum space is then
\begin{equation}
  \label{eq:momspace_int_sl}
  \tilde{V}(p,k) = FT\left[\rho(r;R)V_S(r)\right] +
  g(R)\tilde{\chi}(p;R)\tilde{\chi}(k;R)~,
\end{equation}
where $FT$ represents the Fourier transform and partial-wave projection
shown in Eq.~\eqref{eq:ft_pwp}.

For the semi-locally regulated case, similar to~\cite{Epelbaum2015}, we use
\begin{equation}
  \label{eq:semilocal_reg}
  \rho(r;R) = {\left[1-e^{-{(r/R)}^2}\right]}^4~,
\end{equation}
and
\begin{equation}
  \label{eq:mom_reg}
  \tilde{\chi}(p;R) = e^{-{(pR/2)}^2} = e^{-{(p/\Lambda)}^2}~,
\end{equation}
where $\Lambda\equiv 2/R$. For a brief discussion on the different $\rho(r;R)$
functions used for the locally and semi-locally regulated cases,
see Appendix~\ref{sec:rho_choice}.

\subsubsection{Nonlocal Regulation}\label{sec:nonlocal_reg}

For the fully nonlocal interaction, we take the semi-local
interaction Eq.~\eqref{eq:momspace_int_sl}, including the forms of $\rho(r;R)$ and
$\tilde{\chi}(p;R)$, and modify the first term as follows
\begin{equation}
  \label{eq:momspace_int_nl}
  \tilde{V}(p,k) = \tilde{\chi}(p;R) FT\left[\rho(r;R_<)V_S(r)\right]
\tilde{\chi}(k;R) + g(R)\tilde{\chi}(p;R)\tilde{\chi}(k;R)~.
\end{equation}
The momentum-space regulators multiplying the first term suppress the diagonal
matrix elements where the incoming and outgoing momenta are large but similar,
removing some sensitivity to the choice of $\rho(r;R)$ that we discuss
in~\ref{sec:rho_choice}. The short-distance cutoff used before we take the
Fourier transform, $R_<$, is chosen to be much less than $R$. This allows us to
ensure that the resulting cutoff dependence in the observables is attributable
to the regulator function, $\tilde{\chi}(p;R)$, rather than the Fourier
transform.

\subsection{Two-Body Bound States}\label{sec:two_body_bound_states}
We calculate two-body binding energies by solving the Schr\"odinger equation
\begin{equation}
  \label{eq:shroedinger}
  (\hat{H}_0 + \hat{V})\ket{\psi} = E\ket{\psi}~,
\end{equation}
in coordinate and momentum space. In coordinate space, we tune the counterterm
such that for a desired value $E$, the radial equation
\begin{equation}
  \label{eq:shroedinger_radial}
  -\frac{1}{m}\frac{d^2u}{dr^2} + V(r) u(r) = E\,u(r)~,
\end{equation}
is solved where $u(r)\equiv rR_0(r)$. We have dismissed the centrifugal term as
only s-waves are considered. In momentum space, we
rearrange Eq.~\eqref{eq:shroedinger} such that we have
\begin{equation}
    \label{eq:determinant}
     \hat{G}_0(E)\hat{V}\ket{\psi}  = \ket{\psi}~,
\end{equation}
where $G_0(z)\equiv 1/(z-\hat{H}_0)$. After discretization with the basis states
$|p_i\rangle$, Eq.~\eqref{eq:determinant} becomes an eigenvalue problem that is
easily solved by finding the energies that fulfill
\begin{equation}
  \label{eq:determinant1}
  \det\left[\hat{1}-K_{ij}(E)\right] = 0~,
\end{equation}
where $K_{ij}(E) = \braket{p_i|\hat{G}_0(E)\hat{V}|p_j}$ and we tune the counterterm such
that the requirement Eq.~\eqref{eq:determinant1} is satisfied.

\subsection{Lippmann-Schwinger Equation}\label{sec:lse}
To obtain two-body phase shifts, we solve numerically the
Lippmann-Schwinger Equation for the two-body $t$-matrix
\begin{equation}
\label{eq:lse}
\hat{t} = \hat{V} + \hat{V}\hat{G}_0\hat{t}~.
\end{equation}
In the partial-wave projected momentum basis, considering bosons interacting in
$s$-waves only, we have
\begin{align}
\nonumber
  \braket{p\,|\hat{t}|p^\prime }
  &=\braket{p\,|\hat{V}|p^\prime } + \braket{p\,|\hat{V}\,\hat{G}_0(E+i\epsilon)\,t|p^\prime}~,\\
t(p, p^\prime; E) & = \tilde{V}(p,p^\prime) + \int_0^\infty
  dq\,q^2\,\frac{\tilde{V}(p,q) \,t(q, p^\prime; E)} {E+i\epsilon-q^2/m}
\end{align}
where $m$ is the nucleon mass and $\epsilon \to +0$. From the on-shell matrix element $t(p,p;E=p^2/m)$
we extract the phase shift via
\begin{equation}
  t(p,p;E=p^2/m) = -\frac{2}{m\pi} \frac{1}{p\cot{\delta}-ip}~.
\end{equation}
The scattering length is defined by the effective range expansion
\begin{equation}
  \label{eq:ere}
  p\cot\delta \approx = -\frac{1}{a} + \frac{r_s}{2}p^2~,
\end{equation}
which allows us to calculate it exactly from the on-shell $t$-matrix amplitude
at $p=0$.
\begin{equation}
  a = \frac{m\pi}{2} t(0,0;0)~.
\end{equation}

\subsection{Three-Body Bound States}\label{sec:3b_bound_states}
To calculate three-body binding energies, we start with the equation for a
single Faddeev component of a system containing three identical particles
\begin{equation}
	\label{eq:Faddeev_bound_state}
  \ket{\psi} = \hat{G}_0(E)\hat{t}\hat{P}\ket{\psi}~,
\end{equation}
where
\begin{equation}
	\hat{P} = \hat{P}_{12}\hat{P}_{23} + \hat{P}_{13}\hat{P}_{23}~,
\end{equation}
is the permutation operator with $\hat{P}_{ij}$ interchanging particles
$i$ and $j$~\cite{Gloeckle:99109}.
After projecting onto the partial-wave, momentum basis for three identical
bosons described by two Jacobi momenta $p$ (the relative momentum between
particles 1 and 2) and $q$ (the relative momentum between particle 3 and the
center of mass of the 1--2 subsystem), we discretize the equation and solve for
the bound state energy $E$ using the same techniques as in the two-body case,
as long as $E$ remains below the deepest state in the two-body
spectrum. However, this limitation is in conflict with our goal of
studying the cutoff dependence of two- and three-body observables. As
we go to higher momentum-space cutoffs (smaller $R$ values), spurious
bound states enter the two-body spectrum.  Three-body states quickly
become resonances in this regime, bounded above and below by two-body
bound states. There are two ways that we deal with this.

The first method follows~\cite{Nogga:2005hy} and is repeated here. It involves
\textit{removing} the spurious two-body state from the spectrum by transforming
the potential
\begin{equation}
  \label{eq:state_removal}
  \hat{V} \rightarrow \hat{V} + \ket{\phi}\lambda\bra{\phi}~,
\end{equation}
which takes the eigenvalue of the state $\phi$ and modifies it by an amount
$\lambda$.
Using this transformed potential in the Lippmann-Schwinger equation and taking
the limit
of $\lambda\rightarrow\infty$ (removing the state from the spectrum), we have
\begin{equation}
  \lim_{\lambda\rightarrow\infty} \hat{t}(\lambda) =
  \hat{t} - \ket{\eta}\frac{1}{\braket{\phi|\hat{G}_0|\eta}}\bra{\eta}~,
\end{equation}
as our modified $t$-matrix where
\begin{equation}
	\label{eq:eta}
  \ket{\eta} = \ket{\phi} + \hat{t}\hat{G}_0\ket{\phi}~.
\end{equation}
This only requires that we have the wave function $\braket{p|\phi}$ to
calculate the modified $t$-matrix where that state no longer
contributes a pole. 
In practical calculations using a large, finite $\lambda$ value in
(\ref{eq:state_removal}) is sufficient.
If there are several spurious two-body states, the procedure is repeated
for each of them.

The second method we employ to study the cutoff dependence of
three-body resonances is to look for the resonances in the three-body
phase shifts.

\subsection{Three-Body Phase Shifts}\label{sec:3b_phase_shifts}
In the cutoff regime where spurious two-body bound states exist, we
can scatter a third particle off the spurious deep two-body state
and scan the phase shifts in the energy range between the two-body
states for a resonance. To do this, we calculate the three-body
$T$-matrix using~\cite{Gloeckle:1995jg}
\begin{equation}
\label{eq:T}
  \hat{T} = \hat{t}\hat{P} + \hat{t}\hat{G}_0\hat{P}\hat{T}~,
\end{equation}
which relates to the elastic scattering operator $\hat{U}$ by
\begin{equation}
  \label{eq:U}
  \hat{U} = \hat{P}\hat{G}_0^{-1} + \hat{P}\hat{T}~.
\end{equation}
In the partial-wave-projected, momentum basis, considering bosons interacting
only via $s$-waves, we have
\begin{equation}
	\begin{split}
    \braket{pq|\hat{T}|\phi} & = \braket{pq|\hat{t}\hat{P}|\phi} + \\
  & \int_0^\infty dq^\prime {(q^\prime)}^2 \int_{-1}^1 dx\,
  \frac{t(p,\pi_1,E-3q^2/4m)\,G(q,q^\prime, x)}
  {E+i\epsilon-q^2/m-{(q^\prime)}^2/m-qq^\prime x/m}
   \braket{\pi_2q^\prime|\hat{T}|\phi}~,
	\end{split}
\end{equation}
where the incoming state $\ket{\phi}=\ket{\varphi k}$ contains the
wave function $\varphi(p)$ of the two-body bound state and the relative
momentum $k$ between the third particle and the center of mass of the
two-body subsystem, 
$G(q,q^\prime,x)$ is a geometrical factor introduced by the
permutation operator, $\pi_1 = \sqrt{q^2/4+{(q^\prime)}^2+qq^\prime x}$, and
$\pi_2 = \sqrt{q^2+{(q^\prime)}^2/4+qq^\prime x}$.

The elastic scattering amplitude $M$ is related to the $U$ operator by
\begin{equation}
  M = -\frac{2m\pi}{3}\braket{\phi|\hat{U}|\phi}~,
\end{equation}
and the phase shift by
\begin{equation}
  \label{eq:three-body-pw-amp}
  M = \frac{1}{k\cot\delta-ik}~.
\end{equation}
In the three-body sector, we have a similar effective range expansion
\begin{equation}
  \label{eq:ere-atom-dimer}
  k\cot\delta \approx -\frac{1}{a_{AD}} + \frac{r_{s,AD}}{2}k^2~,
\end{equation}
which defines the atom-dimer scattering length $a_{AD}$ and atom-dimer effective
range $r_{s,AD}$.
We also study the inelasticity parameter given in terms of the $S$-matrix by
\begin{equation}
  \eta = e^{-2\delta_i}~,
  \label{eq:inelasticity}
\end{equation}
where the phase shift is complex and the usual decomposition
\begin{equation}
  \delta = \delta_r + i\delta_i~,
\end{equation}
is taken.

\subsection{Quantitative Uncertainty Analysis}
To analyze the uncertainties induced by short-distance physics of our
regularization procedure, we study in this section the regulator
dependence of observables. Similar to the analysis done by Song
{\textit{et al.}}~\cite{Song:2016ale}, our uncertainty analysis is based on a
simple power series expansion of observables quantities $\mathcal{O}$
of the form
\begin{equation}
  \label{eq:power_series_uncertainties}
  \mathcal{O}(\Lambda) \approx \mathcal{O}_\infty \left[1+\sum_i^\infty
  c_i{\left(\frac{q}{\Lambda}\right)}^i\right]~,
\end{equation}
where $q$ is associated with the low-momentum scale relevant to the calculation;
however, $i$ is \textit{not} assumed to be an integer. For the purposes of this
project, we truncate the summation over $i$ after the first term $i=n$, leaving
\begin{equation}
  \label{eq:lo_correction}
  \mathcal{O}(\Lambda) \approx \mathcal{O}_\infty \left[1 + c_n
  {\left(\frac{q}{\Lambda}\right)}^n\right]~,
\end{equation}
We seek to establish the value of $n$. In Ref.~\cite{Song:2016ale},
$n$ was found by fitting the first few terms in the above expansion
with integer $n$ to the cutoff dependence of observables. Here, we
study the cutoff dependence at very large cutoffs, focus on the
dominant term in the expansion, and fit $n$ itself to data and 
allow for non-integer values.

To extract the power of the leading cutoff correction, we examine both the
$\Lambda$ and the $q$ dependence. The first approach we take to investigate the
$\Lambda$ dependence is to calculate observable $\mathcal{O}$ over a range of
$\Lambda$ values, and fit the results to Eq.~\eqref{eq:lo_correction} for a
range of $n$ values. For each $n$, we evaluate a penalty function that we define
as
\begin{equation}
  \label{eq:penalty_function}
  p_n = \sum_i {\left(\frac{\mathcal{O}_{calc}(\Lambda_i) -
    \mathcal{O}_{fit}(\Lambda_i)}{\mathcal{O}_{calc}(\Lambda_i)}\right)}^2~,
\end{equation}
where $\mathcal{O}_{calc}(\Lambda)$ is the observable calculated for a specific
value of $\Lambda$ and $\mathcal{O}_{fit}(\Lambda)$ is the value of the
observable as it is ``reproduced'' by Eq.~\eqref{eq:lo_correction} and the fit
parameters $\mathcal{O}_\infty$ and $c_n$. Once we have $p_n$ for a range of $n$
values, we search for a minimum $p_n$ where $n$ is optimal.

Griesshammer has shown~\cite{Griesshammer:2015osb} that the $q$ dependence of
observables provides a necessary though insufficient window into the order of
cutoff-dependent corrections. To isolate the $q$ dependence, we have to restrict
the observables we study to those whose $q$ dependence is well understood. Doing
so allows us to calculate the observable at two different cutoffs and study the
relative difference
\begin{equation}
  \label{eq:gh_diff}
  1 - \frac{\mathcal{O}(\Lambda_1)}{\mathcal{O}(\Lambda_2)}
  \approx
  q^n c_n \left[\frac{1}{\Lambda_2^n} - \frac{1}{\Lambda_1^n}\right]~.
\end{equation}
Taking the logarithm, we get
\begin{equation}
  \label{eq:gh_diff_log}
  \ln\left[1 - \frac{\mathcal{O}(\Lambda_1)}{\mathcal{O}(\Lambda_2)}\right] =
  n\ln q + b~,
\end{equation}
where $n$ and $b$ are the slope and intercept that we fit, respectively.

\section{Results}\label{sec:results}

\subsection{Renormalization Group Flow}\label{sec:rgflow}
\begin{figure}
	\includegraphics[width=\linewidth]{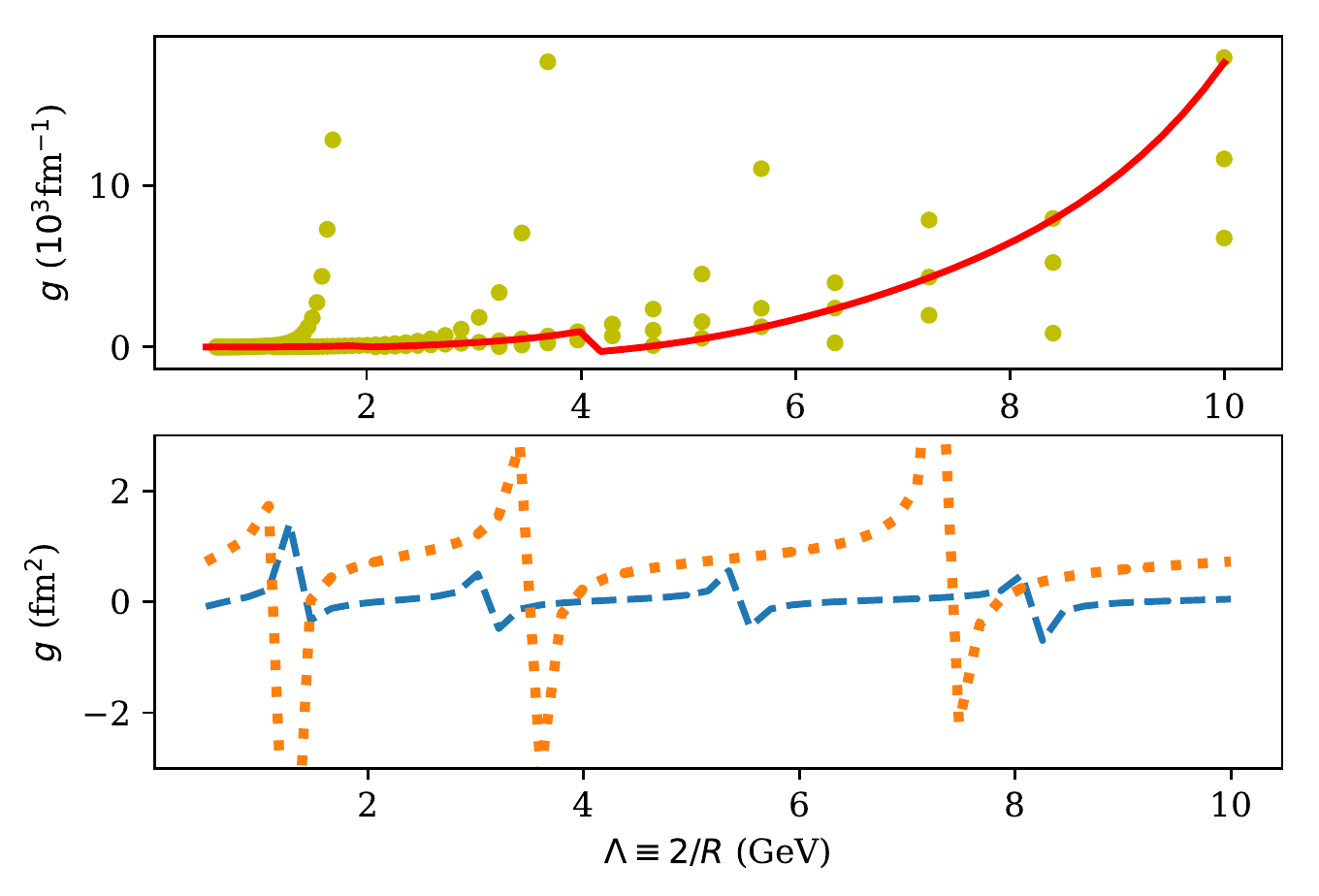}
  \caption{RG flows of the counterterm coupling $g$. The yellow circles in the upper
    plot represent $g(R)$ values calculated with a local regulator and local
    counterterm. The red, solid line in the upper plot are the $g(R)$ values
    used to calculate the phase shifts in
    Fig.~\ref{fig:two_body_phase_shifts}. The blue, dashed line in the lower
    plot corresponds to the semi-locally regulated interaction.  The orange,
    dashed line corresponds to the nonlocally regulated interaction.
  }\label{fig:rg_flows}
\end{figure}

The first thing we compare between the regulation schemes is the RG
flow. We choose to fix the shallowest two-body state at
$B_2 = 2.2$ MeV.  Figure~\ref{fig:rg_flows} shows the stark difference
between the RG flow found using a local counterterm and the RG flows
found with nonlocal counterterms. The main difference is the issue of
uniqueness. For the locally regulated potential, as pointed out
by~\cite{Beane:2000wh}, $g(R)$ has multiple solutions that give a two-body bound
state at the desired binding energy. There is one branch where there exists one
state in the two-body system. Each branch below that branch contains
successively one additional state. The RG flow shown for the locally regulated
interaction connects four of those branches, ``hopping'' downward when it is
easier to add an additional state than to continue to maintain the shallowness
of the fixed state. Only two of the ``hops'' are visible in the plot due to the
scale and the relative difference between the magnitudes of $g$ between the
different branches. Note also the difference in the units of the upper and
lowers plots if Fig.~\ref{fig:rg_flows}. There is a factor of $R^3$ that comes
from the Fourier transform and partial-wave projection of $\chi(r;R)$.

The other two functions shown in the lower plot of Fig.~\ref{fig:rg_flows}
are qualitatively very
similar. They correspond to the semi-local and nonlocal regulation schemes.
While the same $\rho(r;R)$ is used in both, the prescription is somewhat
different as one can see from Eq.~\eqref{eq:momspace_int_sl}
and Eq.~\eqref{eq:momspace_int_nl}. The
semi-local regulation scheme brings in spurious bound states faster than the
nonlocal regulation scheme, but as mentioned before, nonlocal regulation cuts
off the potential at large incoming and outgoing momenta, suppressing
high-momentum contributions.  Still, they are very similar interactions, thus
they provide very similar RG flows.

\subsection{Two-Body Scattering}\label{sec:two-body-scattering}
\begin{figure}
	\includegraphics[width=\linewidth]{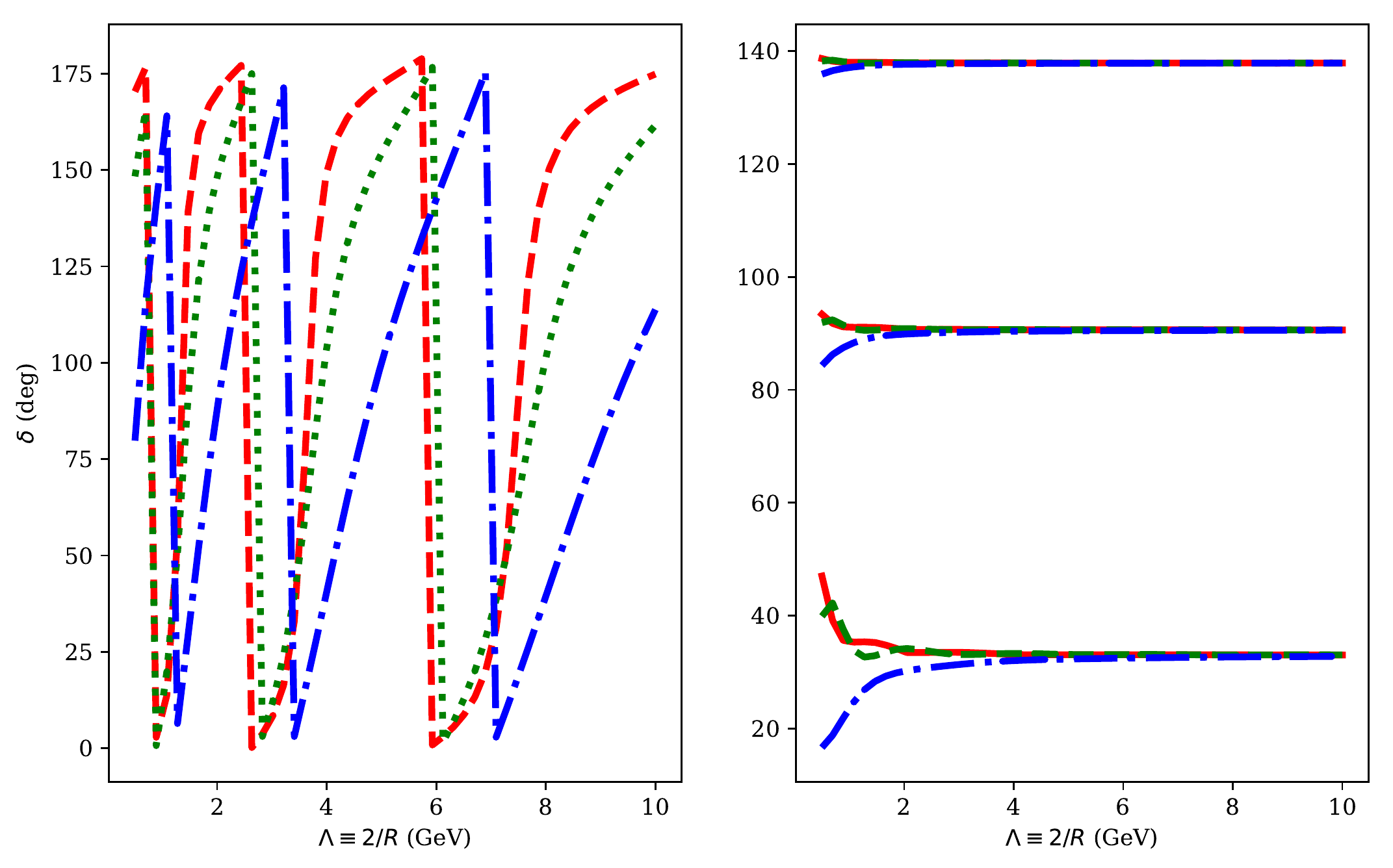}
  \caption{[Left] Cutoff dependence of the s-wave phase shifts at $E = 1$ (red,
    dashed), $10$ (green, dotted), and $100$ MeV (blue, dot-dashed) calculated
    via a nonlocally regulated potential without a counterterm.  [Right] Cutoff
    dependence of the s-wave phase shifts at (from top to bottom) $E = 1$, $10$,
    and $100$ MeV in the center-of-mass frame. The solid, red lines are the phase
    shifts calculated from a locally regulated potential. The green, dashed
    lines are the phase shifts at the same energies calculated with a
    semi-locally regulated interaction. The blue, dot-dashed lines are the phase
    shifts using a nonlocally regulated interaction. All three schemes include a
    contact-like counterterm.
  }\label{fig:two_body_phase_shifts}
\end{figure}

As the different regulation schemes are tuned to reproduce the same
shallow state at $B_2=2.2$ MeV, we expect that differences in
low-energy scattering observables are highly suppressed when large
cutoffs are employed. We calculate the phase shifts using all three
regulation schemes and show the results in
Fig.~\ref{fig:two_body_phase_shifts}. The left plot contains the phase
shifts of an non-renormalized, nonlocally regulated potential with
$g(R) = 0$, demonstrating the strong cutoff dependence of low-energy
observables and the need for a counterterm. The most important feature
of the right plot is the agreement between the different regulation
schemes. 
It is also worth mentioning the ``turning point'' $\Lambda$ value at which phase
shifts clearly begin to flatten out.
At low energies, the point is near 2 GeV.
As the scattering energy increases that point increases as well.
Importantly, this behavior agrees with studies of the OPE
potential~\cite{Song:2016ale,Nogga:2005hy} where similar convergence behavior is
found across a range of partial-wave channels.
Our $C_3$ value is chosen to mimic the OPE in the bosonic sector such that we
can expect similar renormalization behavior.
Observing this similarity is consistent with the known result that the
one-pion-exchange potential goes like an inverse cube potential at short
distances (high cutoffs)~\cite{Sprung:1994,PavonValderrama:2005gu}.

\begin{figure}
  \includegraphics[width=0.9\linewidth]{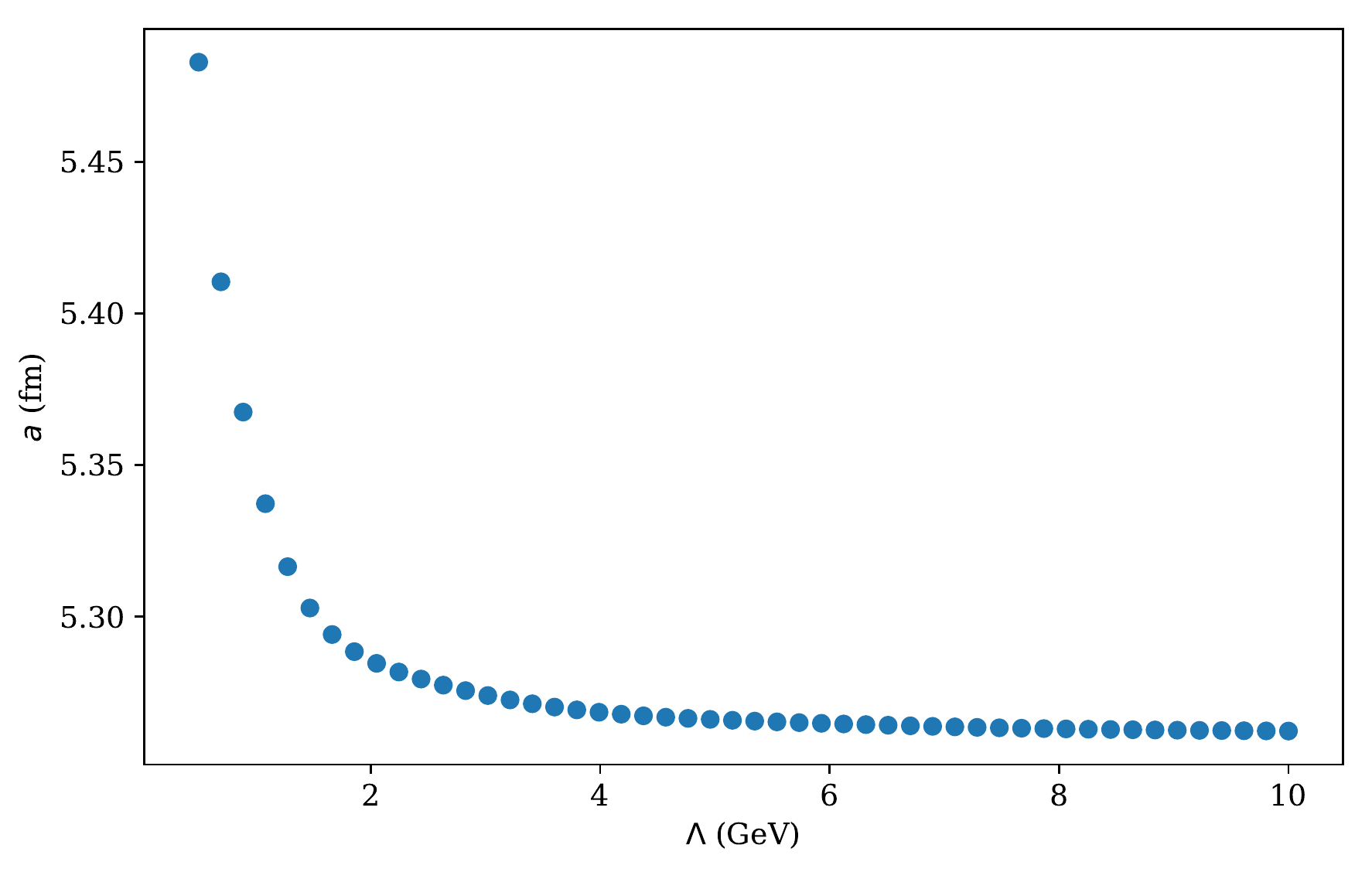}
  \caption{The scattering length is shown as a function of the high-momentum
  (short-distance) cutoff. The blue circles are the numerical results.
  }\label{fig:two_body_scattering_length}
\end{figure}

It is clear from Fig.~\ref{fig:two_body_phase_shifts} that a two-body
contact interaction is sufficient to renormalize the two-body phase
shifts. The corresponding result for the two-body scattering length is
shown in Fig.~\ref{fig:two_body_scattering_length}.

\begin{figure}
  \includegraphics[width=\linewidth]{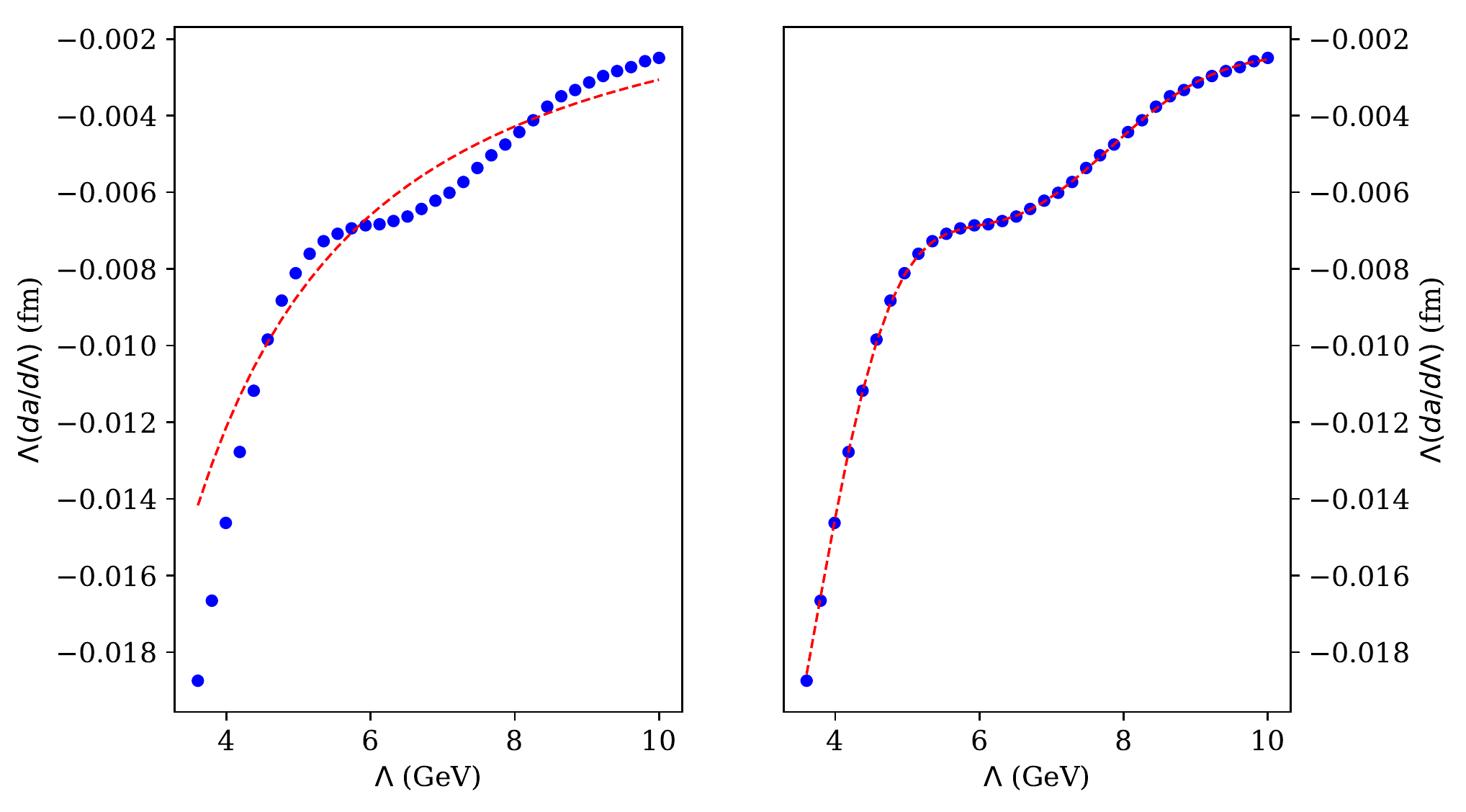}
  \caption{[Left] RG analysis of the two-body scattering length as a function of
    the cutoff. Blue circles represent the data. The red, dashed line represents
    a fit to Eq.~\eqref{eq:lo_correction} with $n=1.5$. [Right] The blue circles
    represent the same data as the left-hand plot. The red, dashed line
    represents a fit to Eq.~\eqref{eq:rgi_correction} with $n_{\min}=1.7$.
  }\label{fig:a_analysis}
\end{figure}

One of the advertised, key advantages of EFT is quantifiable
uncertainty which in turn requires a power counting that orders
contributions in the Hamiltonian according to their
importance. These uncertainties have usually two sources: (i) the
truncation of the low-energy expansion and (ii) uncertainties that are
introduced when low-energy counterterms are fitted to data. Here we
focus on the first source of uncertainties and some information on
this truncation error is contained in the convergence behavior of
observables as the short-distance cutoff is increased. To study this
problem, we first chose a range of cutoffs over which to fit the
scattering length to Eq.~\eqref{eq:lo_correction}. However, as the window of
cutoffs over which the fit was carried out was narrowed to include only the highest
values of $\Lambda$, the resulting $n$ was found to be unstable.
As a result, we plotted $\Lambda(da/d\Lambda)$, shown in
Fig.~\ref{fig:a_analysis}.
The solid, red line in the left-hand plot of Fig.~\ref{fig:a_analysis} is the
expected $\Lambda(da/d\Lambda)$ dependence based on a fit to
Eq.~\eqref{eq:lo_correction} with $n=1.5$.
Clearly, there is behavior in the cutoff dependence of the scattering length
that is not captured by the simple form assumed in
Eq.~\eqref{eq:lo_correction}.

Empirically, we model the residual cutoff dependence by
\begin{equation}
  \label{eq:rgi_correction}
  \Lambda\frac{da}{d\Lambda} \approx \frac{1}{\Lambda^n}\left[A +
  B\cos \left({h\Lambda^{1/3}+f}\right)\right]~,
\end{equation}
where $A, B, h$ and $f$ are treated as fit parameters.
We choose a range of $n$ values over which we carry out the fit and evaluate the
quality of the fit with Eq.~\eqref{eq:penalty_function} at each value.
The right-hand plot of Fig.~\ref{fig:a_analysis}
shows $\Lambda(da/d\Lambda)$ in blue circles with $n_{\min}=1.7$.
The red, dashed line in the left-hand plot of Fig.~\ref{fig:a_analysis}
represents Eq.~\eqref{eq:rgi_correction} with the fit parameters found when
using $n_{min}$.
The agreement between the data and the empirical formula is excellent.

\begin{figure}
  \includegraphics[width=\linewidth]{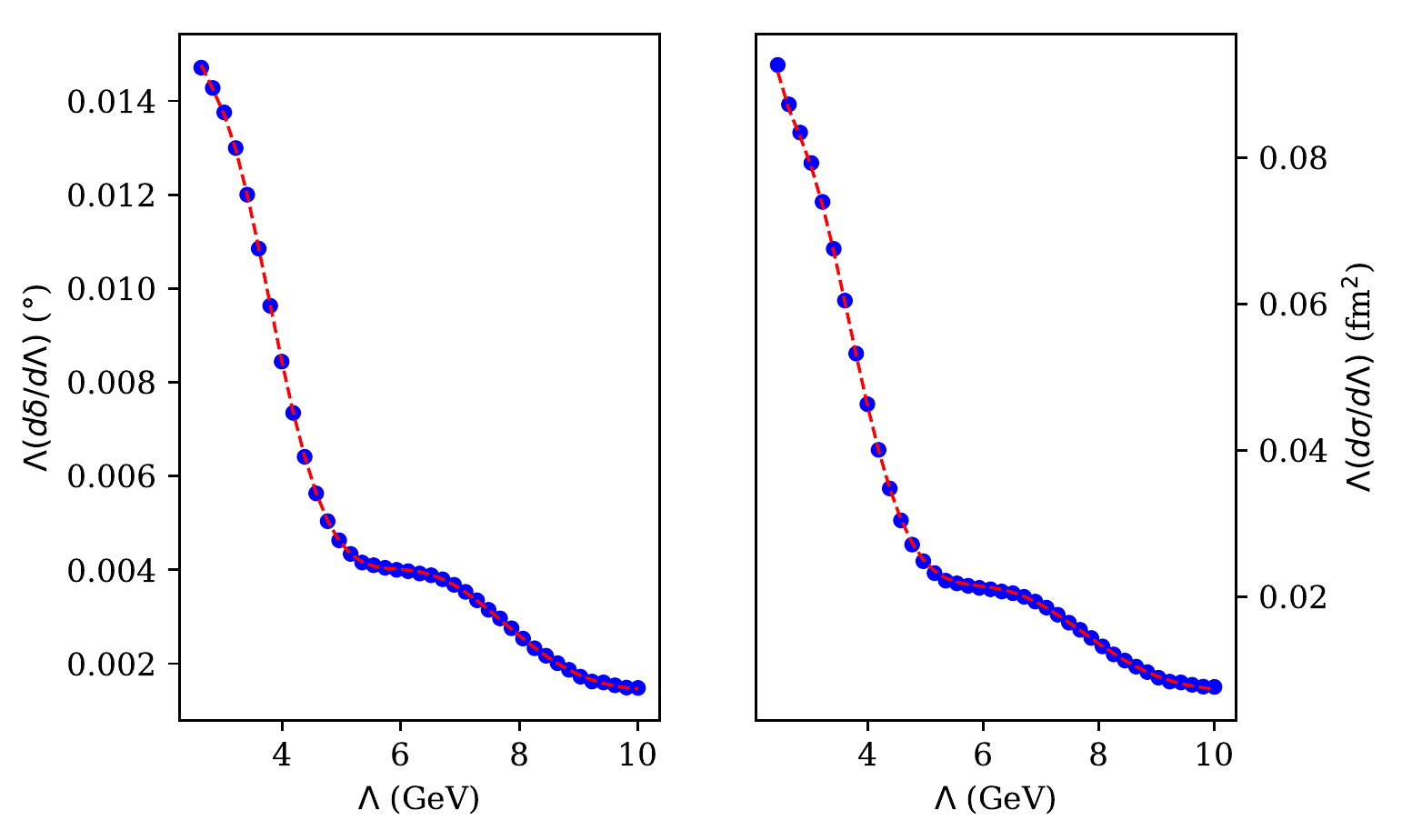}
  \caption{[Left] RG analysis of the phase shift at a center-of-mass momentum of
    106 MeV  as a function of the cube root of the cutoff. Blue circles
    represent the data. The red, dashed line represents a fit to
    Eq.~\eqref{eq:rgi_correction} with $n_{\min}\approx 1.7$. [Right] The same analysis
    of the cross section at a center-of-mass momentum of 106 MeV as a function
    of the cube root of the cutoff. The legend is the same as in the left-hand
    plot.}\label{fig:delta_sigma_analysis}
\end{figure}

We expect that all low-energy, two-body observables come with similar cutoff
dependence.
In keeping with our study of the cutoff dependence of the scattering length, we
applied the same analysis to the phase shifts and cross sections.
In Fig.~\ref{fig:delta_sigma_analysis} we plot the results.
In both cases, the calculation was performed at a relative, center-of-mass
momentum of 106~MeV.
The analyses produced minima of the penalty functions
(Eq.~\eqref{eq:penalty_function}) near $n_{\min}=1.7$.
Similar analyses performed at different energies produced similar results.
The only trend worth mentioning is the slight decrease of $n_{\min}$ to
approximately 1.5 as the scattering energy increases.
Overall, the agreement between the data and Eq.~\eqref{eq:rgi_correction} found
for the scattering length is found for the phase shift and cross section as
well.
The $n_{\min}$ values are collected in Table~\ref{tab:nmins}.

Interestingly, the $h$ values vary by less than a few percent around 1.5
MeV$^{-1/3}$ between the observables.
This fairly constant oscillation frequency matches up with the frequency of new
bound states in the RG flow.
As shown below, this correspondence carries over to the three-body sector as
well.

\begin{figure}[ht]
  \includegraphics[width=\linewidth]{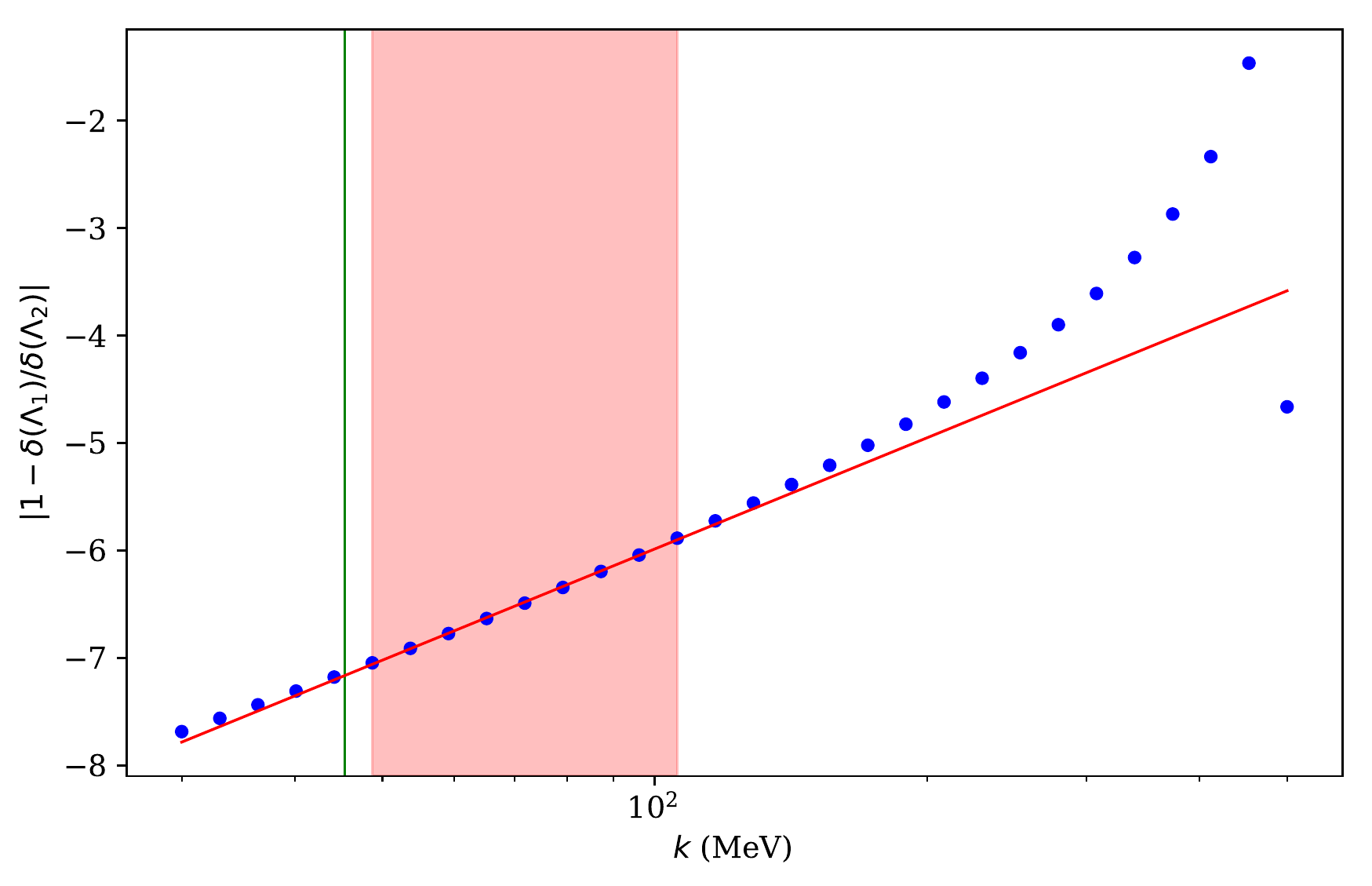}
  \caption{Residual cutoff corrections to the two-body phase shifts as a
    function of the relative momentum.
    The blue circles represent the numerical calculation.
    The red line represents a fit to Eq.~\eqref{eq:gh_diff_log}, resulting in
    $n=1.5$.
    The pink, shaded region represents the range of $k$ over which the fit was
    performed.
    The vertical, green line is the binding momentum $\gamma$.
  }\label{fig:delta_gh_analysis}
\end{figure}

The order of corrections is independent of the method used to obtain it.
In that spirit, we apply in addition to our modified power series expansion the
method proposed by Griesshammer~\cite{Griesshammer:2015osb}.
Fig.~\ref{fig:delta_gh_analysis} shows the comparison of the phase shifts at
$\Lambda=3408$ and $6704$ MeV.
By Eq.~\eqref{eq:gh_diff_log}, we expect the behavior to be linear.
In fact, we are able to extract a reliable slope of $n=1.5$ by fitting the data
to Eq.~\ref{eq:gh_diff_log}.
Unfortunately, we found that other observables such as the cross section and
$k\cot\delta$ provide unreliable results.
Specifically, zeros and unpredictable crossings precluded the extraction of
linear behavior.
Selecting the phase shifts as the quantities of interest follows naturally from
these unfortunate conditions as discussed by
Griesshammer~\cite{Griesshammer:2015osb}.

\subsection{Three-Body Scattering}

\begin{figure}[ht]
	\includegraphics[width=0.9\linewidth]{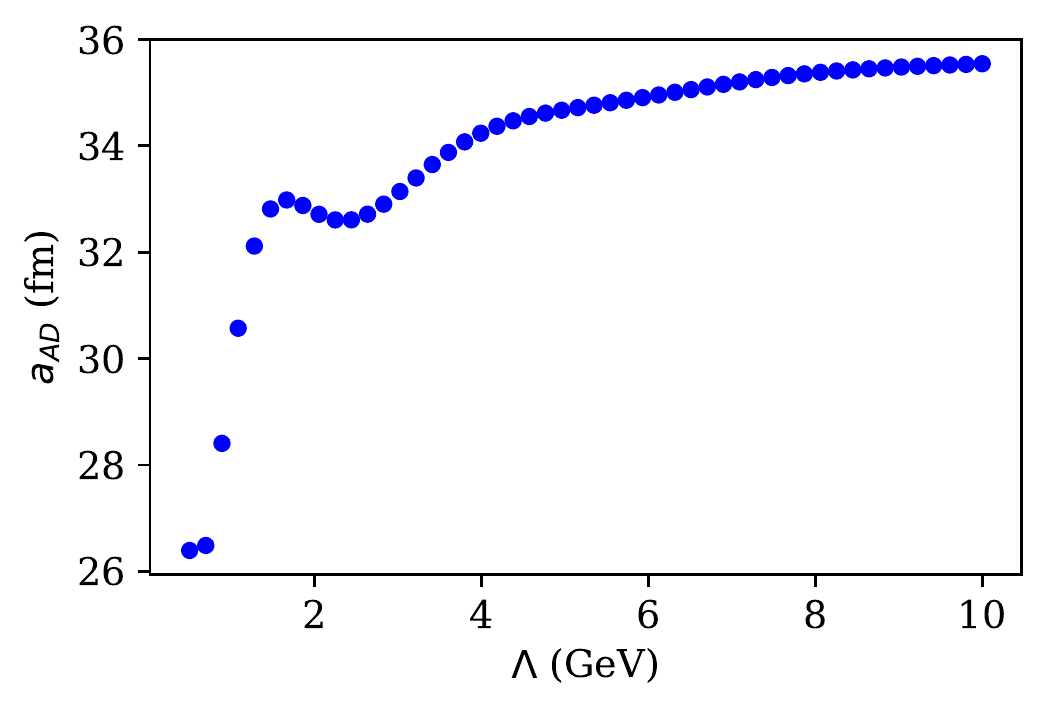}
  \caption{The cutoff dependence of the atom-dimer scattering length.
  }\label{fig:a_AD_cutoff_dependence}
\end{figure}

The first observable in the three-body sector that we study is the atom-dimer
scattering length.
Figure~\ref{fig:a_AD_cutoff_dependence} shows the convergence of $a_{\textrm{AD}}$
with respect to the momentum cutoff $\Lambda$, clearly demonstrating that a
two-body contact term is sufficient to renormalize three-body observables.

\begin{figure}[ht]
  \includegraphics[width=0.9\linewidth]{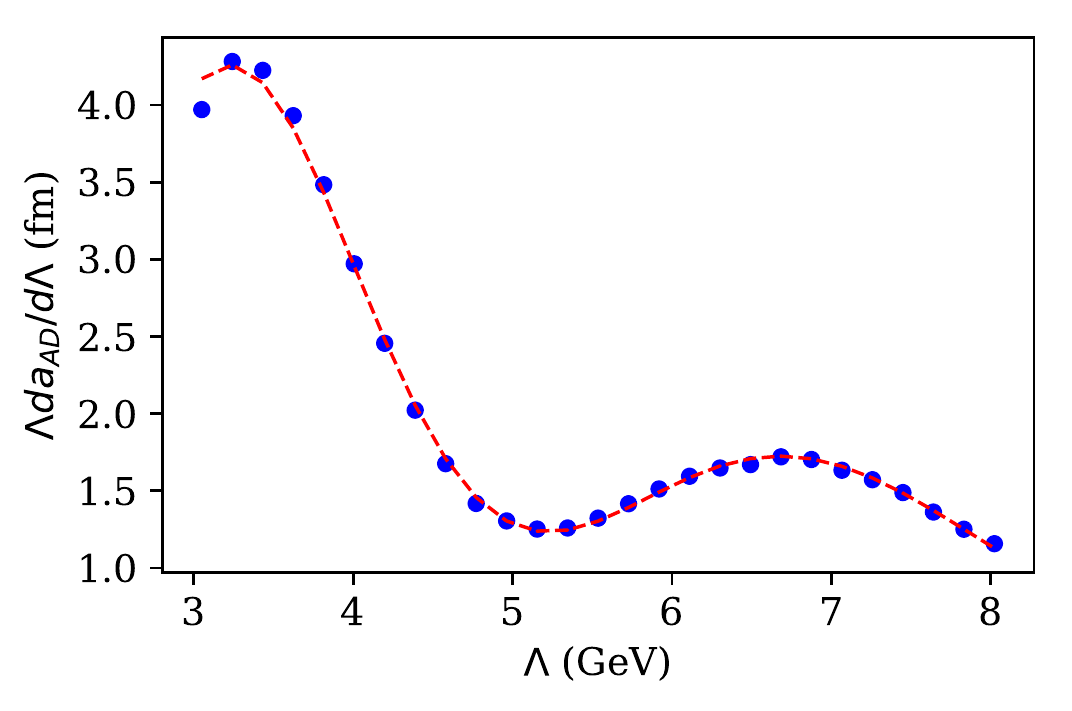}
  \caption{$\Lambda (da_{AD}/d\Lambda)$ as a function of the momentum-space
    cutoff. The blue circles are the calculation. The solid, red line is the
    fit to Eq.~\eqref{eq:rgi_correction} with $n_{\min}=1.3$.
  }\label{fig:a_AD_analysis}
\end{figure}

Again, we apply the analysis based on Eq.~\eqref{eq:rgi_correction} to the
atom-dimer scattering length.
The results are shown in Fig.~\ref{fig:a_AD_analysis}.
As in the two-body sector, Eq.~\eqref{eq:rgi_correction} is able to accurately
describe the oscillatory convergence behavior occuring on top of the typically
expected $\Lambda$-dependence.
The fit was performed over a range of cutoffs --- from
$\Lambda_\textrm{lower}=3.1$~GeV to $\Lambda_\textrm{upper}=8.1$~GeV.
For the atom-dimer scattering length, the best fit to
Eq.~\eqref{eq:rgi_correction} occurs at $n_{\min}=1.3$.
Because this analysis involves the derivative of the observable with respect to
$\Lambda$ and three-body observables are particularly difficult to obtain to
arbitrary accuracy, we are often forced to constrain our fit window.
The atom-dimer scattering length, as well as the other three-body observables
presented below, are selected because they provide stable results over a
significant range of cutoffs.

\begin{figure}[ht]
  \includegraphics[width=\linewidth]{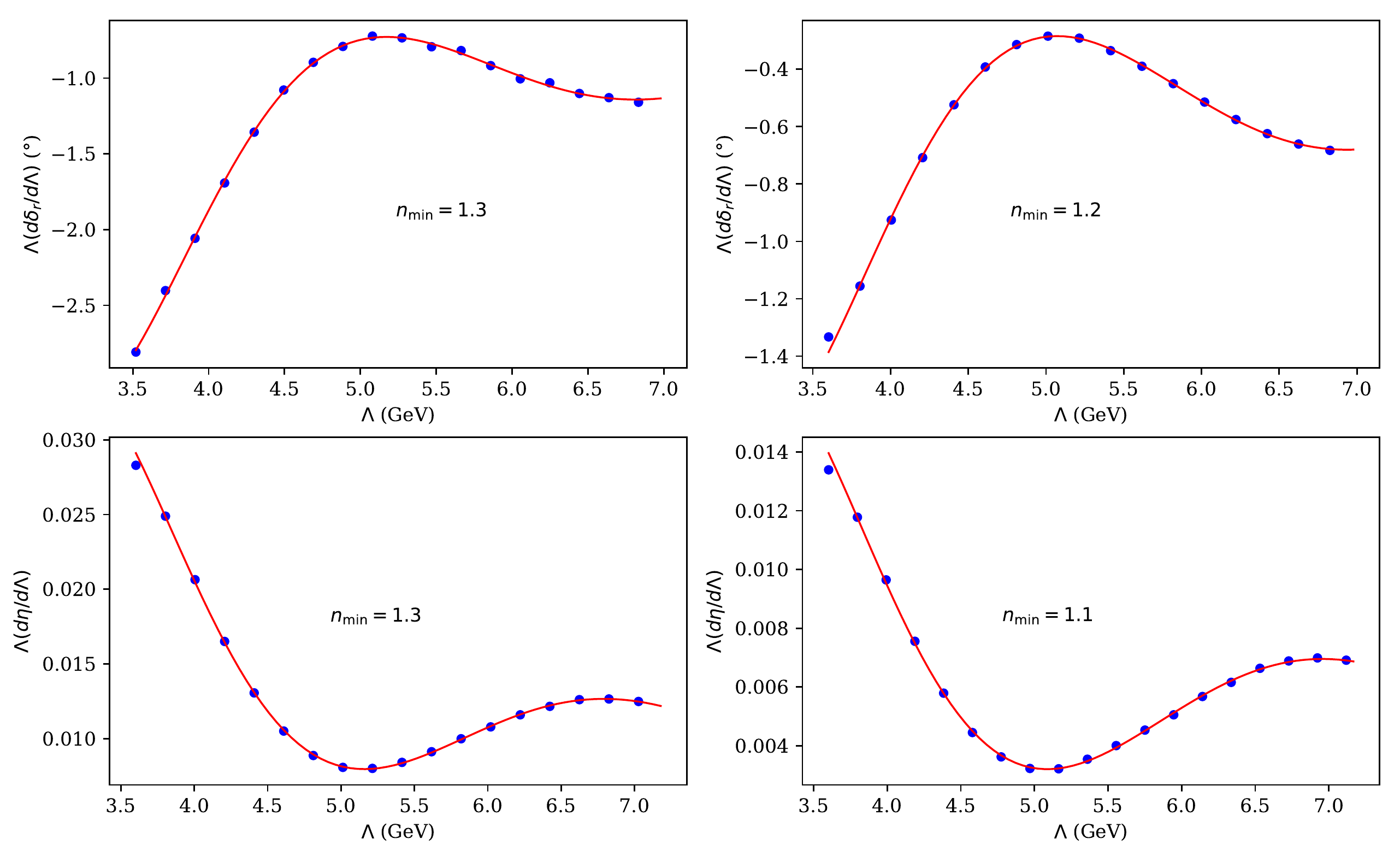}
  \caption{[Upper Left] Eq.~\ref{eq:rgi_correction}-based analysis of the 2+1
    phase shift at $E=10$~MeV. The blue circles are the calculation. The solid,
    red line is the fit to Eq.~\ref{eq:rgi_correction}. [Upper Right] Same
    analysis and legend applied to the 2+1 phase shift at $E=50$~MeV. [Lower
    Left] Inelasticity at $E=50$~MeV.  [Lower Right] Inelasticity at
    $E=100$~MeV.
  }\label{fig:3b_scatter_analysis}
\end{figure}

In addition to the atom-dimer scattering length, we also conduct analyses of
three-body phase shifts and inelasticities at center-of-mass, kinetic energies
of 10, 50, and 100~MeV.
The results are shown in Fig.~\ref{fig:3b_scatter_analysis}.
The $n_{\min}$ values, ranging from 1.1 to 1.3, used to plot the solid, red
lines corresponding to Eq.~\ref{eq:rgi_correction} are tabulated in
Table~\ref{tab:nmins}.
The bounds of the cutoff range are included as well to assure the reader that
the behavior represents a significant and relevant portion of the cutoff
dependence.

\subsection{Three-Body Bound States}\label{sec:three-body-bound}

\begin{figure}[ht]
	\includegraphics[width=\linewidth]{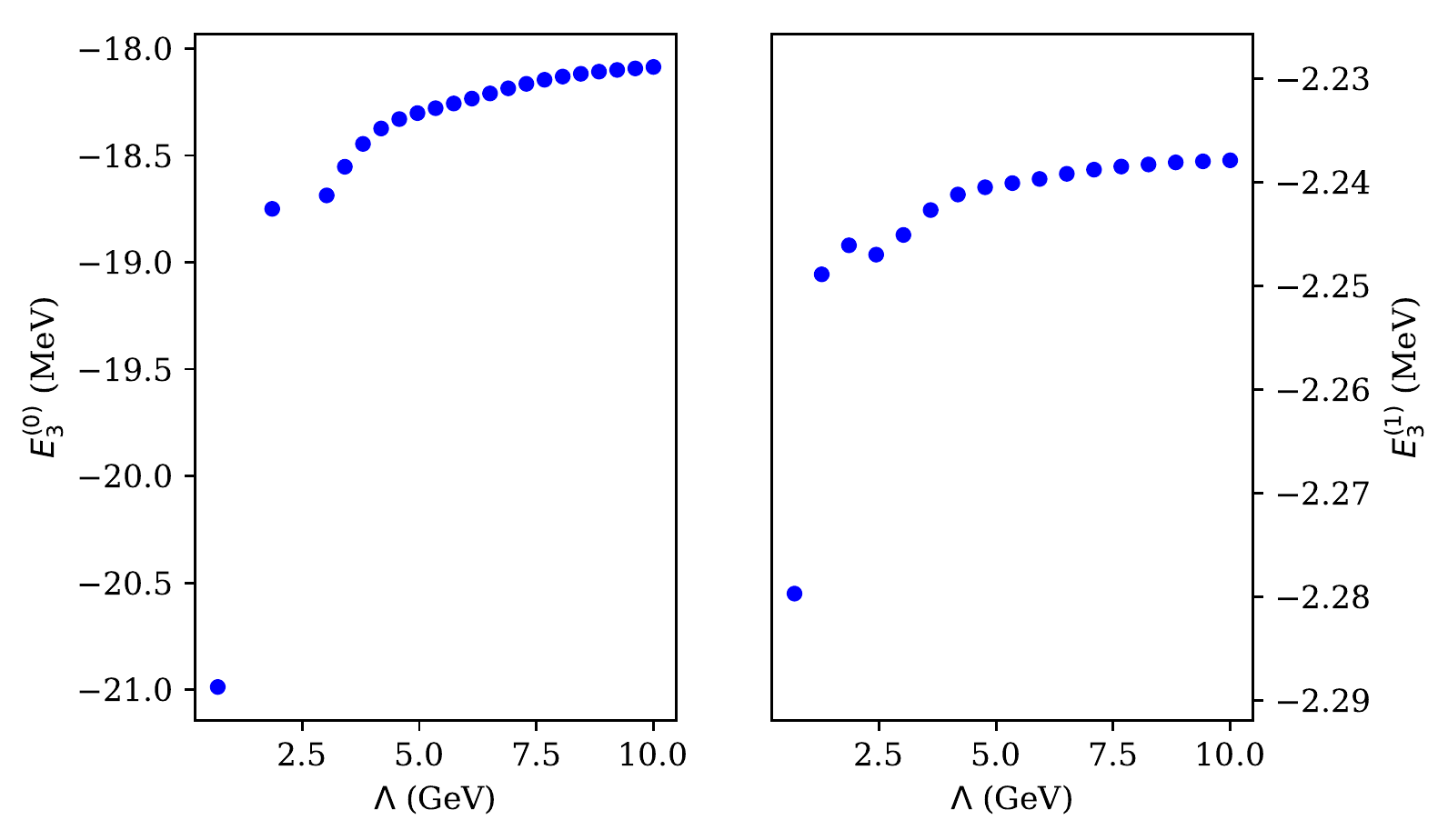}
  \caption{[Left] Three-body ground state/resonance energy as a function of the
    short-distance cutoff. [Right] Three-body excited state/resonance energy as
    a function of the short-distance cutoff.
  }\label{fig:three_body_states}
\end{figure}

One of the main goals of these efforts has been to examine the
sufficiency of a two-body counterterm to renormalize three-body
observables.
In Fig.~\ref{fig:three_body_states} we plot the cutoff dependence of the
three-body binding energies associated with two three-body states that appear in
the system defined by the nonlocally regulated interaction
Eq.~\eqref{eq:momspace_int_nl}.
The results shown come from the solution of Eq.~\eqref{eq:Faddeev_bound_state},
though equivalent results were found by calculating the three-body phase shifts
defined by Eq.~\eqref{eq:three-body-pw-amp} and scanning for resonances.
The ground state and excited state binding energies at $\Lambda=10$~GeV are
-18.086~MeV and -2.2379~MeV, respectively.
The primary feature of Fig.~\ref{fig:three_body_states} is the convergence of
the binding energies in the infinite $\Lambda$ limit.
At $\approx 2$~GeV, the binding energies (or rather, the resonant energies)
begin to flatten out, just as in the two-body phase shifts.

\begin{figure}[ht]
	\includegraphics[width=\linewidth]{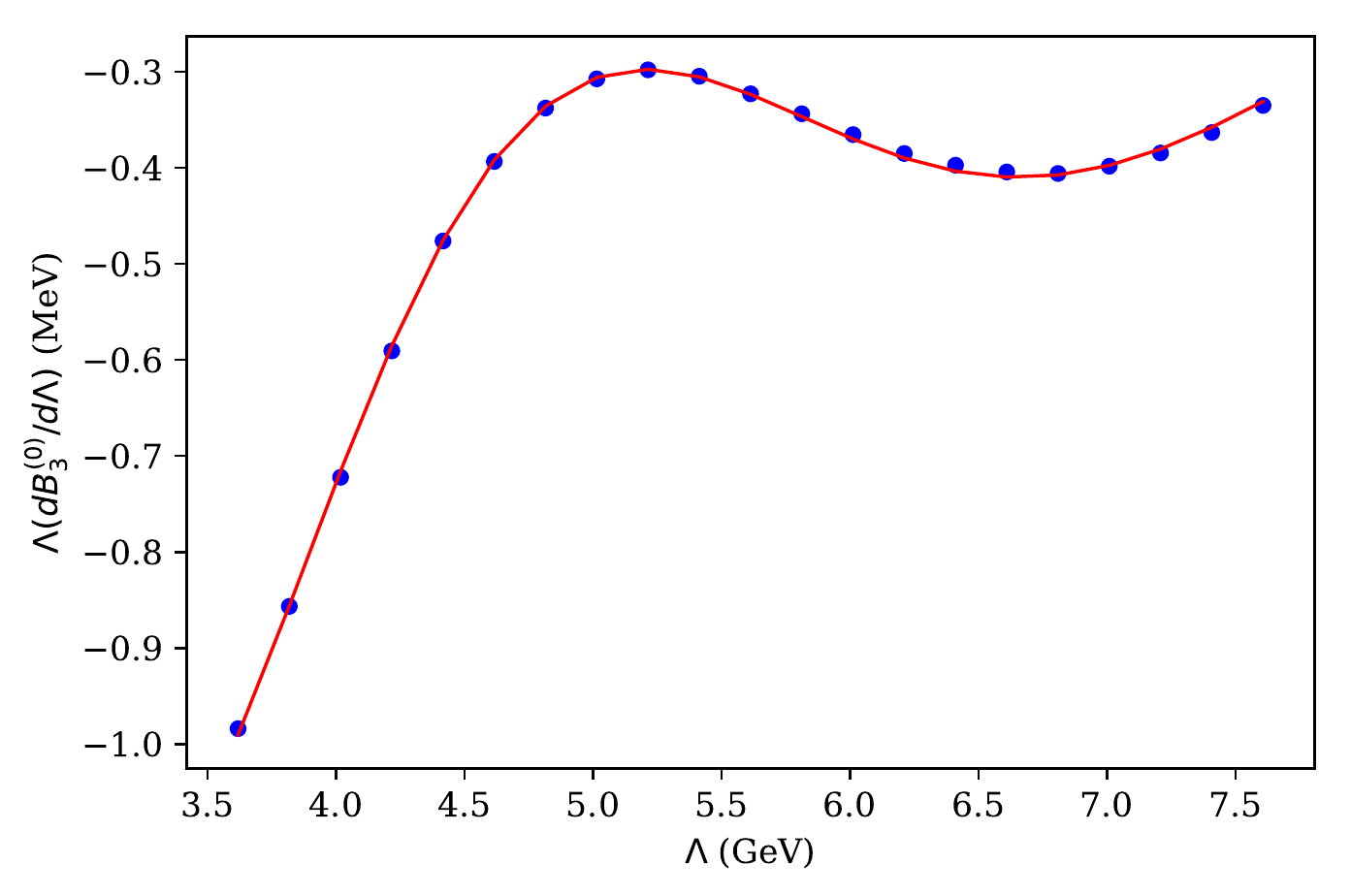}
  \caption{RG analysis of the three-body, ground-state binding energy. The blue
    circles are the calculation. The red line represents a fit to
    Eq.~\eqref{eq:rgi_correction} with $n_{\min}=1.4$ and the values of $h$ and
    $f$ taken from the same fit of the atom-dimer scattering length.
  }\label{fig:3b_gs_cutoff_dep}
\end{figure}

Unfortunately, small inaccuracies in the three-body binding energies left only
small windows of cutoffs over which a fit to Eq.~\eqref{eq:rgi_correction} could
be performed when all four fit parameters were treated as such.
Using the values of $h$ and $f$ from the fit of the atom-dimer scattering length to
Eq.~\eqref{eq:rgi_correction}, we fit only $A$ and $B$ for the ground state
binding energy and show the results in Fig.~\ref{fig:3b_gs_cutoff_dep}.
An $n_{\min}$ value of 1.4 is found to minimize the penalty function, and 
the form of Eq.~\eqref{eq:rgi_correction} is further validated.

Throughout all of the three-body observables, we see a consistency among the
$h$ values.
Notably, it is enforced manually for the three-body ground state.
They range from 1.4 to 1.5 MeV$^{-1/3}$ which is also consistent with the $h$
values found by fitting the two-body observables.
This consistency between the two- and three-body sectors can be seen in
Table~\ref{tab:nmins} which establishes the pervasive nature of these oscillations.

\setlength\tabcolsep{12pt}
\begin{center}
  \begin{table}
    \begin{tabular} { c c c c c c }
    \hline\hline
    Observable & $n_{\min}$ & $\Lambda_{\textrm{lower}}$ (GeV) &
    $\Lambda_{\textrm{upper}}$ (GeV) & h (MeV$^{-1/3}$) \\
    \hline
    $a(\Lambda)$ & 1.7 & 3.6 & 10.0 & 1.5 \\
    $\delta(\Lambda;E=12\textrm{MeV})$ & 1.7 & 2.6 & 10.0 & 1.5 \\
    $\sigma(\Lambda;E=12\textrm{MeV})$ & 1.7 & 2.4 & 10.0 & 1.5 \\
    $\delta(k)$ & 1.5 & 3.4 & 6.7 & --- \\
    $a_{AD}(\Lambda)$ & 1.3 & 3.1 & 8.1 & 1.5 \\
    $\delta_{2+1}(\Lambda;E=10\textrm{MeV})$ & 1.3 & 3.7 & 7.7 & 1.4 \\
    $\delta_{2+1}(\Lambda;E=50\textrm{MeV})$ & 1.2 & 3.7 & 7.0 & 1.4 \\
    $\eta_{2+1}(\Lambda;E=50\textrm{MeV})$ & 1.3 & 3.7 & 7.0 & 1.5 \\
    $\eta_{2+1}(\Lambda;E=100\textrm{MeV})$ & 1.1 & 3.7 & 7.1 & 1.4 \\
    $E_3^{(0)}$ & 1.4 & 3.5 & 7.8 & 1.5* \\
    \hline
  \end{tabular}
  \caption{$n_{\min}$ values for various two- and three-body observables
  alongside the bounds of cutoffs over which the fit to
  Eq.~\eqref{eq:rgi_correction} was performed as well as the frequency that
  optimizes the fit. * The $h$ value for $E_3^{(0)}$ was taken from the fit of
  $a_{AD}$.}\label{tab:nmins}
\end{table}
\end{center}

\section{Summary}
\label{sec:summary}

In this manuscript, we have set out to understand the renormalization
properties of the FRIC potential in the two- and
three-body sector.
In particular, we have studied the regulator dependence of observables such as
two-body phase shifts, three-body binding energies, the atom-dimer scattering
length, phase shifts, and inelasticity parameter.
Motivated by a recent development in the nuclear theory community, we did these
calculations using different, frequently used regulator functions.

Our results in the two-body sector confirm that the two-body sector is
properly renormalized. One input parameter is required (at leading
order) to renormalize one low-energy counterterm and thereby the
two-body sector. In the three-body sector, we have demonstrated that a
three-body force is not needed at leading order to renormalize
three-body observables for the inverse cube interaction.

In both the two- and three-body sectors, we have observed significant
oscillatory behavior in the cutoff dependence of observables. These
oscillations are not captured by a simple power series expansion.

Instead, we have empirically found that a generalized oscillatory dependence of
the form presented in Eq.~\eqref{eq:rgi_correction} allows accurate fits of the
data to be made and a much clearer picture of the power of the cutoff dependence
to be revealed.

Our analysis strongly indicates that $n$ is smaller in the three-body sector
than in the two-body sector.
This would suggest that a three-body force is needed at next-to-leading order.

Our analysis also indicates that $n$ is consistent with approximately 1.5
for two-body observables and approximately 1 for three-body
observables. It is an interesting question whether this has any
significance for the counting of two- and three-body counterterms in
an EFT for the inverse cube potential. For example, the singular
$1/r^2$ has been considered previously as the starting point for an
EFT expansion in Ref.~\cite{Long:2007vp}, however the inverse cube and
all other singular coordinate space potentials need their own
independent analysis.

Having tested several different local, semi-local, and nonlocal regulators and
having found no significant differences above $\approx$2~GeV, we conclude that
these oscillations are most likely attributable to the singular nature of the
inverse cube potential in coordinate space.

In the future, we plan to carry out an analysis of higher order corrections in
the three-boson and three-nucleon sector.
However, we plan to also extend our work to the infinite range inverse cube
potential that is of relevance to the atomic dipole interaction.
This will let us combine the results obtained by M\"uller~\cite{Mueller:2013}
with three-body observables and study the dependence of three-body observables
on the boundary condition employed in the two-body sector.
A more detailed analysis of the short-distance behaviour of the three-nucleon
wave function might also provide novel insights into the power counting of
electroweak currents~\cite{Valderrama:2014vra}.

\appendix
\section{Local Regulator Sensitivity}\label{sec:rho_choice}
\begin{figure}[ht]
  \includegraphics[width=\linewidth]{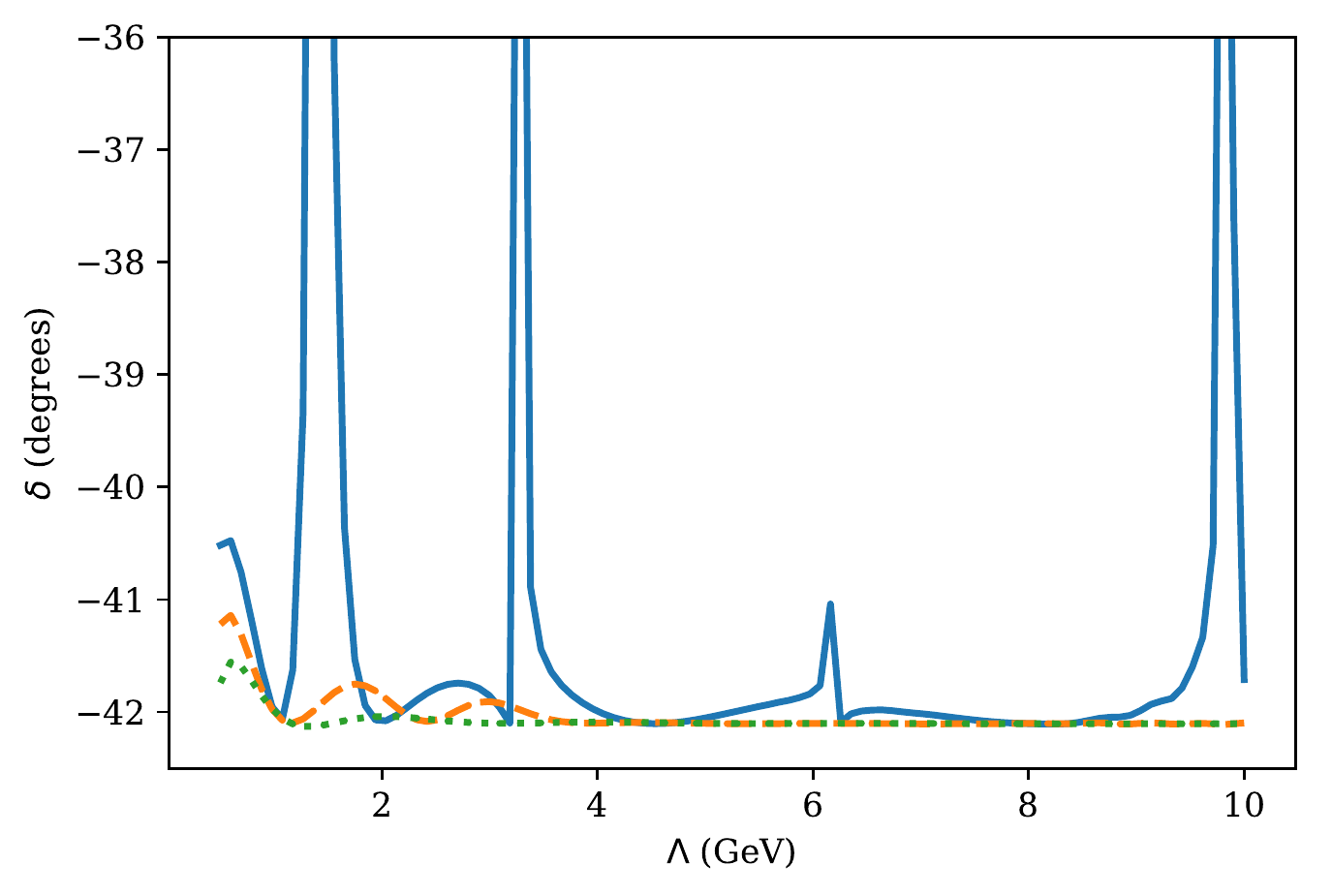}
  \caption{Phase shifts at a center-of-mass energy of 1 MeV for three different
    regulators. Per Eq.~\eqref{eq:gen_local_reg}, the solid, blue line corresponds
    to $n_1=3$ and $n_2=1$. The dashed, yellow line corresponds to $n_1=4$ and
    $n_2=1$. The dotted, green line corresponds to $n_1=2$ and $n_2=4$.
  }\label{fig:local_reg_compare}
\end{figure}

To regulate the interaction
\begin{equation}
  V_S(r) = -C_3 \frac{e^{-m_\pi r}}{r^3}~,
\end{equation}
a general (local) regulator, $\rho(r;R)$, can be used such that the limit
\begin{equation}
  \label{eq:regular_condition}
  \lim_{r\rightarrow 0} \rho(r;R)V_S(r)~,
\end{equation}
is finite.
We use regulators of the form
\begin{equation}
  \label{eq:gen_local_reg}
  \rho(r;R) = {(1-e^{-{(r/R)}^{n_1}})}^{n_2}~,
\end{equation}
whose small $r$ behavior goes like $r^{n_1n_2}$.
As long as $n_1n_2 \ge 3$, the regulator sufficiently meets the requirement of
Eq.~\ref{eq:regular_condition}.
However, our earliest calculations using the semi-local regulation scheme with $n_1=3$
and $n_2=1$ gave inconsistent results.
Specifically, we observed unexpected cutoff dependence in the phase shifts as
shown in Fig.~\ref{fig:local_reg_compare}.
Simply increasing $n_2$ to 4 such that $n_1n_2 = 4 > 3$ removes the dramatic
changes in the phase shift.
We have also compared our local regulators with those used by
others~\cite{Epelbaum2015,Binder:2018pgl}.
In the interest of consistency and to ensure we avoid unexpected cutoff
dependence, we have used a local regulator of the form $n_1=2$ and $n_2=4$ for
the calculations carried out it in this work.
The unexpected cutoff dependence was observed exclusively when using semi-local
regulation and only when $n_1n_2=3$.

\begin{acknowledgements}
  This work has been supported by the National Science Foundation
  under Grant No. PHY-1555030, and by the Office of Nuclear Physics,
  U.S. Department of Energy under Contract No. DE-AC05-00OR22725.
  A.D. acknowledges the support  by the Alexander von Humboldt Foundation
  under Grant No. LTU-1185721-HFST-E.
\end{acknowledgements}

\bibliographystyle{spphys}       

\end{document}